\documentclass[trackchanges]{aastex701}

\begin{document}

\title{Updated Mutual Inclination Measurement for 14\,Her\,b and c\footnote{Data and code are available in \url{https://github.com/gyxiaotdli/mini_Agatha/tree/main/HD145675}}}

\author[orcid=0000-0001-6753-4611]{Guang-Yao Xiao}
\affiliation{State Key Laboratory of Dark Matter Physics, Tsung-Dao Lee Institute \& School of Physics and Astronomy, Shanghai Jiao Tong University, Shanghai 201210, China}
\email[show]{gyxiao\_tdli@sjtu.edu.cn}  

\author[]{Fabo Feng} 
\affiliation{State Key Laboratory of Dark Matter Physics, Tsung-Dao Lee Institute \& School of Physics and Astronomy, Shanghai Jiao Tong University, Shanghai 201210, China}
\email[show]{ffeng@sjtu.edu.cn}

\begin{abstract}
The mutual inclination $\psi$ between orbits within multi-planetary systems is difficult to measure directly. Through a joint analysis of RVs, Hipparcos-Gaia absolute astrometry and relative astrometry of JWST, \citet{Bardalez2025} recently found two possible values of $\psi_{bc}={32_{-15.1}^{+13.6}}^{\circ}$ and $\psi_{bc}={145.0_{-11.1}^{+15.8}}^{\circ}$ for the nearby, non-transit planet system, 14\,Her. By incorporating additional astrometry from Gaia second data release (DR2), we have further constrained the orbital orientation of 14\,Her\,b, resulting in a definitive and unambiguous mutual inclination of $\psi_{bc}={35.3_{-7.3}^{+6.8}}^{\circ}$. The unusual misaligned orbital architecture of 14\,Her system may serve as a benchmark for dynamical studies.  
\end{abstract}



\section{Introduction} 
14\,Her (HD\,145675) is nearby, bright and metal-rich star that has been monitored by several radial velocity (RV) surveys for about three decades. 
Based on observations from the W. M. Keck Observatory, \citet{Butler2003ApJ} reported a super-Jupiter, 14\,Her\,b, in a moderately eccentric, thousand-day orbit. 
Subsequent RV follow-up observations revealed an additional giant planet, 14\,Her\,c, in a wider separation \citep{Wittenmyer2007ApJ, Rosenthal2021ApJS}, sparking interest in their orbital architecture and dynamical history.

The mutual inclination between planets is a key indicator of the dynamical evolution of multi-planetary systems but is challenging to measure directly, especially in non-transiting systems due to limited constraints on orbital orientation. Recent studies have combined different techniques (mainly RV and astrometry) to determine the true orbital configurations of these systems,  
revealing significant misalignment in a few cases (e.g., \citealt{Xuan2020MNRAS, An2025AJ, Zhang2025AJ}). 

For the 14\,Her system, \citet{Benedict2023AJ} combined Hubble Space Telescope (HST) Fine Guidance Sensor data and long-term RVs to determine the planet orbits, finding a mutual inclination of $\psi_{bc}=62^{\circ}\pm12^{\circ}$ between b and c, which significantly deviates from coplanarity. 
By incorporating absolute astrometry from the Hipparcos and Gaia DR3 (also known as proper motion anomaly, PMA), \citet{Bardalez21} also found a misaligned orbit with $\psi_{bc}={96.3_{-29.1}^{+36.8}}^{\circ}$. However, it seems that their analysis suffered from inclination (or longitude of the ascending node) degeneracy \citep{Feng2025MNRAS}, leading to two possible sets of orbits and thus leaving the true architecture of the system uncertain. This issue still exists even with the inclusion of additional relative astrometry from JWST. 
Recently, \citet{Bardalez2025} directly imaged 14\,Her\,c using JWST/NIRCam coronagraphy in F444W band and reported bimodal mutual inclinations of $\psi_{bc}={32_{-15.1}^{+13.6}}^{\circ}$ and $\psi_{bc}={145.0_{-11.1}^{+15.8}}^{\circ}$, based on joint analysis of archival RVs, JWST relative astrometry JWST, and Hipparcos-Gaia DR3 absolute astrometry.  

To resolve the degeneracy, we optimized our analysis by incorporating multiple Gaia data releases (DR2 and DR3) and simulating the Gaia epoch data with Gaia Observation Forecast Tool \footnote{\url{https://gaia.esac.esa.int/gost/index.jsp}} (GOST). Compared to other methods (e.g., \texttt{orvara},\citealt{Brandt2021AJ}), the additional astrometry from Gaia DR2 is helpful for determining the true orientation of 14\,Her\,b. This method currently serves as a temporary solution until Gaia DR4 becomes available.
Further details are presented in our previous works \citep{Feng2023MNRAS, Xiao2024MNRAS}. 

\section{Analysis and Results}
The RVs of 14\,Her are from the ELODIE \citep{Naef2004}, KECK \citep{Rosenthal2021ApJS}, HET \citep{Benedict2023AJ}, and APF \citep{Vogt2014PASP}, while astrometry data include Hipparcos \citep{vanLeeuwen2007}, Gaia DR2 and DR3 absolute astrometry \citep{GaiaCollaboration2018,GaiaCollaboration2023}, and direct imaging data of JWST \citep{Bardalez2025}. The primary fitting parameters of our method include the orbital period $P$, RV semi-amplitude $K$, eccentricity $e$, argument of periastron $\omega$ of stellar reflex motion, orbital inclination $i$, longitude of ascending node $\Omega$, mean anomaly $M_{0}$ at the minimum epoch of RV data and five astrometric offsets ($\Delta \alpha*$, $\Delta \delta$, $\Delta \varpi$, $\Delta \mu_{\alpha*}$ and $\Delta \mu_\delta$) of barycenter relative to GDR3. The semi-major axis $a$ of the planet relative to the host, the mass of planet $m_{\rm p}$, and the epoch of periastron passage $T_{\rm p}$ can be derived from above orbital elements. Most priors are uniform (see Table 1 of \citealt{Xiao2024MNRAS}). The stellar mass, $0.98\pm0.04\,M_{\odot}$ \citep{Bardalez21}, is assigned a Gaussian prior. 

The posterior sampling is carried out with the parallel-tempering Markov Chain Monte Carlo (MCMC) sampler \texttt{ptemcee} \citep{Vousden2016}. We employ 30 temperatures, 100 walkers, and 80,000 steps per chain to generate posterior distributions for all the fitting parameters, with the first 20,000 steps being discarded as burn-in. For each chain, the initial values of $i$, $\Omega$, $\Delta \alpha*$, $\Delta \delta$, $\Delta \mu_{\alpha*}$ and $\Delta \mu_\delta$ are drawn from uniform distributions, in order to explore the potential bimodal distributions. 

\begin{figure*}
    \centering
	\includegraphics[width=0.99\textwidth]{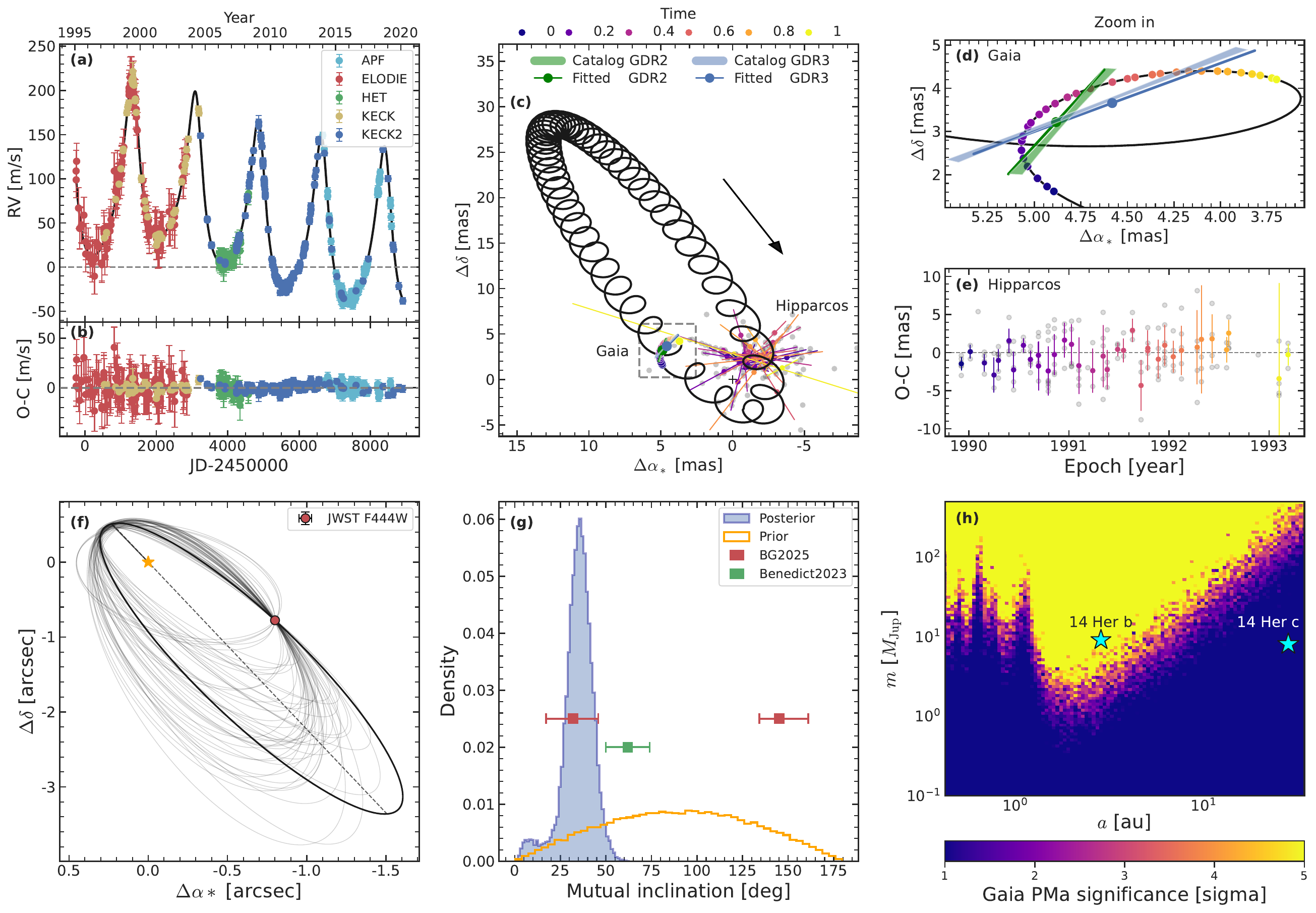}
    \caption{\textbf{Panel (a)}: Best-fit to RVs of various instruments. \textbf{Panel (b)}: RV residuals. \textbf{Panel (c)}: Best-fit to Hipparcos, Gaia DR2 and DR3 astrometry. The black thick line denotes the optimal orbit of 14\,Her, while the arrow indicates its direction of motion. \textbf{Panel (d)}: Zoom in on the rectangle region of panel (c) where depicts the best fit to Gaia GOST data. \textbf{Panel (e)}: Residuals of Hipparcos abscissa. The brightness of points gradually increases with observation time. \textbf{Panel (f)}: Best-fit to JWST data. 50 orbits are randomly drawn from the MCMC chain. \textbf{Panel (g)}: Posterior distribution of $\psi$ between the orbits of 14\,Her\,b and c, compared to the prior, \citet{Benedict2023AJ} (Benedict2023) and \citet{Bardalez2025} (BG2025). \textbf{Panel (g)}: PMa significance map between Gaia DR2 and DR3. The short-scale astrometric variation induced by 14\,Her\,b is very significant.}
    \label{fig:five_p}
\end{figure*}

Figure\,\ref{fig:five_p} presents the primary results of our joint fitting. For 14\,Her\,b, the best-fit orbital parameters are $a_{b}={2.84}_{-0.04}^{+0.04}\,\rm au$, $e_{b}=0.371_{-0.003}^{+0.003}$, and $i_{b}={147.3_{-2.7}^{+2.2}}^{\circ}$, in great agreement with the values found by \citet{Bardalez2025}. We note $i_b$ is differ by $\pi$ with the result $i_{b}={35.7\pm3.2}^{\circ}$ of \citet{Benedict2023AJ}, which may be due to the use of different coordinate system convention or insufficient posterior sampling \citep{Feng2025MNRAS}.
The inclination only shows a single peak. For 14\,Her\,c, the best-fit orbital parameters are $a_{c}={28.1}_{-6.8}^{+6.4}\,\rm au$, $e_{c}=0.64_{-0.10}^{+0.06}$, and $i_{c}={111.9_{-4.5}^{+5.4}}^{\circ}$, also consistent with those reported by \citet{Bardalez2025}. The large uncertainties for $a_c$ and $e_c$ are due to the incomplete phase coverage of RVs.
The final calculated mutual inclination is $\psi_{bc}={35.3_{-7.3}^{+6.8}}^{\circ}$, comparable with the result $\psi_{bc}={32_{-15.1}^{+13.6}}^{\circ}$ of \citet{Bardalez2025}. Therefore, both 14\,Her\,b and c have the same direction of orbital motion. This unambiguous measurement of $\psi_{bc}$ might serve for further detailed dynamical investigations. 

\begin{acknowledgments}
This work is supported by the National Natural Science Foundation of China (NSFC) under Grant No. 12473066.
\end{acknowledgments}

%



\bibliography{sample701}{}
\bibliographystyle{aasjournalv7}



\end{document}